\documentclass[aps,prl,twocolumn,showpacs]{revtex4}

\usepackage{times}

\begin{document}

\title{Quantum information cannot be completely hidden in correlations:\\
implications for the black-hole information paradox}

\author{Samuel L.\ Braunstein$^\dagger$ and Arun K.\ Pati$^\ast$}
\affiliation{$^\dagger$Computer Science, University of York, 
 York YO10 5DD, United Kingdom}
\affiliation{$^\ast$Institute of Physics, Sainik School Post, 
Bhubaneswar-751005, Orissa, India}

\begin{abstract}
The black-hole information paradox has fueled a fascinating effort 
to reconcile the predictions of general relativity and those of 
quantum mechanics. Gravitational considerations teach us that black 
holes must trap everything that falls into them. Quantum mechanically 
the mass of a black hole leaks away as featureless (Hawking) radiation, 
but if the black hole vanishes, where is the information about the 
matter that made it? We treat the states of the in-fallen matter 
quantum mechanically and show that the black-hole information paradox 
becomes more severe. Our formulation of the paradox rules out 
one of the most conservative resolutions: that the state of the 
in-falling matter might be {\it hidden\/} in correlations between 
semi-classical Hawking radiation and the internal states of the black 
hole. As a consequence, either unitarity or Hawking's semi-classical 
predictions must break down. Any resolution of the black-hole information 
crisis must elucidate one of these possibilities.
\end{abstract}


\maketitle

In 1917 Vernam invented his one-time pad cipher. In its simplest form, 
the cipher encodes a message using a random key to determine whether or 
not to flip each message bit. The original message may be retrieved from 
the encoded form by anyone who has access to the encoded message as well 
as the (secret) key. Because the encoded message still contains unflipped 
bits one might worry whether portions of the original message can be 
extracted from it. This concern was put to rest by Shannon 
when he proved that the encoded bit string contained no information 
of the original message \cite{Shannon} --- it was indistinguishable from 
a random bit string.  Where then does the information reside? It is 
neither in the encoded message nor is it in the key. Instead, all the
information has been transformed into pure correlations between these two
strings. How does this result apply to the black-hole information paradox?
In fact, this classical result has fueled the conjecture that while 
black-hole information cannot strictly be found within the Hawking 
radiation \cite{swh0}, it can nonetheless be hidden within correlations 
between that radiation and something else \cite{page,thoft}.

A direct quantum analogue to the one-time pad would encode an arbitrary 
quantum state into the correlations between two subsystems, with 
{\it none\/} of the information about that state accessible from either 
subsystem alone. Interestingly, such a quantum analogue is 
impossible for any {\it pure-state\/} encoding into two 
subsystems \cite{secret}. For example, for the mapping
\begin{equation}
\alpha|0\rangle+\beta|1\rangle \rightarrow
\alpha(|00\rangle-|11\rangle)+\beta(|01\rangle+|10\rangle) \;,
\end{equation}
at least some information about $\alpha$ and $\beta$ can be gleaned
by looking at either of the two final subsystems alone \cite{secret}.
That this holds generally is particularly surprising since the one-time 
pad is often cited as the classical analogue of quantum 
teleportation \cite{Collins}. In quantum teleportation, Alice is given 
an arbitrary quantum state whose details are unknown to her. In addition, 
she shares one-half of an entangled state with Bob. Alice is allowed to 
send any classical message to Bob after which he is to reconstruct the 
original state. Like the one-time pad, the shared entangled state 
(analogous to the secret key) contains no information about the original 
state. Similarly, the message Alice sends Bob (analogous to the one-time pad
encoded message) contains no information about the original state. 
Notwithstanding this close analogy, its impossibility indicates that 
something must be missing from the above description. In fact, we shall see
that a full description of quantum teleportation contains a third subsystem 
(an `environment') that allows Alice to decohere her measurements thus 
yielding a classical message.

Consider now an arbitrary quantum state (mixed or potentially entangled 
to some external reference state) which is encoded into a larger 
Hilbert space through some unitary process. Suppose this encoding process 
completely hides the information about that state from a particular 
subsystem of that Hilbert space (i.e., the state of that subsystem shows 
no dependence on the state being being hidden). We prove that the hidden 
information is wholly encoded in the remainder of Hilbert space with no 
information stored in the correlations between the two subsystems 
\cite{secret}. Put differently, we prove that, unlike classical 
information, quantum mechanics allows only one way to completely hide 
an arbitrary quantum state from one of its subsystems: by moving it to 
its other subsystems. More importantly, we prove that this result is 
robust to imperfections in the hiding process. 
We call this the ``no-hiding theorem.''

The no-hiding theorem sheds new light on the black-hole 
information paradox and accentuates the crisis for quantum physics. In
particular, it has been speculated that at least some of the information 
that falls into a black hole (encoded by matter and radiation) may be 
found in the correlations between the Hawking radiation leaving the black 
hole and the black hole's internal state or late-epoch radiation
\cite{page,thoft,wil,schiffer93,pres94,kraus,bose,obadia}.
So long as Hawking's semi-classical characterization of the black-hole 
radiation \cite{swh0} is accurate, we prove that the quantum 
information about the in-fallen matter cannot be hidden in these correlations.

{\bf Perfect hiding processes:}
Consider a process which takes an arbitrary input state $\rho_I$ from 
subspace $I$ and encodes it into a larger Hilbert space. This will be a 
hiding process if there exists some subspace $O$ (the output) whose state
$\sigma_O$ has no dependence on the input state. In other words, our hiding 
process maps $\rho_I\rightarrow \sigma_O$ with $\sigma$ fixed for all $\rho$.
The remainder of the encoded Hilbert space may be regarded as an ancilla $A$.
Thus, the entire system may be represented in terms of two subsystems $O$ and 
$A$.  Now for this process to be physical, it must be linear and unitary. By 
linearity, it is sufficient to study the action on an arbitrary pure state 
$\rho_I=|\psi\rangle_I\,{}_I\langle\psi|$. Unitarity allows us to suitably
enlarge the ancilla so that the hiding process can be represented 
as a mapping from pure states to pure states. The hiding process can now be 
expressed in terms of the Schmidt decomposition of the final state
\begin{equation}
|\psi\rangle_I
\rightarrow \sum_{k=1}^K
\sqrt{p_k}\,|k \rangle_O \otimes|A_k(\psi)\rangle_A \;. \label{Xform}
\end{equation}
Here $p_k$ are the $K$ non-zero eigenvalues of $\sigma$, $\{|k\rangle\}$ are
its eigenvectors, and both $\{|k\rangle\}$ and the ancilla states
$\{|A_k\rangle\}$ are orthonormal sets.

In Eq.~(\ref{Xform}) we have explicitly allowed for a possible dependence 
of the ancilla states on $|\psi\rangle$. However, the physical nature of this 
hiding process places a restriction on this dependence. By linearity 
the ancilla will consist of an orthonormal set of states even for
a superposition of inputs
$|A_k(\alpha|\psi\rangle+\beta|\psi_\perp\rangle)\rangle
=\alpha|A_k(\psi)\rangle + \beta|A_k(\psi_\perp)\rangle,$ 
where $|\psi_\perp\rangle$ denotes any state orthogonal to $|\psi\rangle$.
Taking the inner product between two such ancilla states yields 
\begin{equation}
\alpha^\ast\beta \,\langle A_l(\psi)|A_k(\psi_\perp)\rangle
+\beta^\ast\alpha \,\langle A_l(\psi_\perp)|A_k(\psi)\rangle =0 \;.
\end{equation}
Thus, for arbitrary complex values of $\alpha$ and 
$\beta$, all cross-terms above must vanish. 
Given any orthonormal basis $\{|\psi_j\rangle,\; j=1,\ldots,d\}$ spanning 
the input states we may now define an orthonormal set of states, 
$|A_{kj}\rangle \equiv |A_k(\psi_j)\rangle$, spanning a $Kd$-dimensional 
Hilbert space that completely describes the reduced state of the ancilla.
Unitarity allows us to map any orthonormal set into any 
other. Thus, we are free to write these as 
$|A_{kj}\rangle = |q_k\rangle\otimes|\psi_j\rangle\oplus 0$ 
where $\{|q_k\rangle\}$ is an orthonormal set of $K$ states and 
$\oplus\, 0$ means we pad any unused dimensions of the ancilla space
by zero vectors. Under this mapping we see that the arbitrary
input states $|\psi\rangle$ are completely encoded within
the ancilla and Eq.~(\ref{Xform}) becomes
\begin{equation}
|\psi\rangle_I \rightarrow
\sum_k\sqrt{p_k}\, |k\rangle_O\otimes ( |q_k\rangle 
\otimes|\psi\rangle \oplus 0)_A \;. \label{thm1}
\end{equation}
Since we may swap $|\psi\rangle$ with any other state in the ancilla
using purely ancilla-local operations, we conclude that any 
information about $|\psi\rangle$ that is encoded globally is in fact 
encoded entirely within the ancilla. No information 
about $|\psi\rangle$ is encoded in system-ancilla correlations (nor, in fact,
in system-system correlations).

{\bf Imperfect hiding processes:}
Unlike perfect hiding processes, for which we found it sufficient
to consider pure input states, imperfect hiding must allow for some
imprecision in the encoding. To fully specify the mapping, we now need
to describe its action on entangled states; this further guarantees that
the mapping is completely positive and therefore physical. 

If the input 
subsystem $I$ is initially entangled with an (external) reference 
subsystem $I'$ in state 
$|\psi\rangle_{I'I}\equiv \sum_j\sqrt{\lambda_j}|j',j\rangle_{I'I}$ then
linearity and Eq.~(\ref{thm1}) imply that a perfect hiding process on
an entangled state has the form
\begin{eqnarray}
|\psi\rangle_{I'I} &\rightarrow& |\Psi^{\rm perfect}\rangle_{I'OA} \\
&\equiv& \sum_{jk}\sqrt{\lambda_jp_k}\, |j'\rangle_{I'}\otimes
|k\rangle_O\otimes ( |q_k\rangle \otimes|j\rangle \oplus 0)_A 
\nonumber \;,
\end{eqnarray}
i.e., the specification we sought takes the form
$\rho_{I'I}\equiv|\psi\rangle_{I'I}\;{}_{I'I}\langle \psi|
\rightarrow \rho_{I'}\otimes \sigma_O$, where $\rho_{I'}$ is the reduced 
state of the reference. An imperfect process can be described more 
generally by $\rho_{I'I}\rightarrow \rho_{I'O}$ where the output only 
imprecisely hides the input with
\begin{equation}
{\rm tr}\; |\rho_{I'O}- \rho_{I'}\otimes \sigma_O|
 < \epsilon \label{criterion} \;,
\end{equation}
for some $\epsilon$. The choice of trace norm is most appropriate since 
it places a bound on the probability for any observable to distinguish 
these states \cite{NC}. We can now use the fidelity to quantify the 
overlap between the global description of imperfectly hidden states and 
the perfect form given in Eq.~(\ref{thm1}).  Since the fidelity satisfies 
$F(\rho,\sigma) \ge 1-\frac{1}{2}{\rm tr}\; |\rho-\sigma|$, we have
\begin{equation}
F(\rho_{I'O},\, \rho_{I'}\otimes \sigma_O) \ge 1-\epsilon/2 \;.
\end{equation}
By definition, the fidelity is the maximum overlap over all purifications
of the pair of states. Equivalently, we may fix one purification and 
maximize the overlap based on varying the other purification \cite{Schu}.
Let us choose the obvious purification of $\rho_{I'O}$, namely, the actual 
global output which we denote $|\Psi^{\rm imperfect}\rangle$.
Further, the tensor product $\rho_{I'}\otimes\sigma_O$ is highly 
restrictive and it is easy to see that {\it any\/} purification thereof
must take the 
form of $|\Psi^{\rm perfect}\rangle$ (up to some unitary operation
on the ancilla). As a consequence, the global state of 
the imperfect output will strongly overlap with some global state whose 
form perfectly satisfies the no-hiding theorem
\begin{equation}
\langle \Psi^{\rm imperfect} |\, \Psi^{\rm perfect} \rangle
\ge 1-\epsilon/2 \label{overlap} \;,
\end{equation}
or, stated differently, 
\begin{equation}
|\Psi^{\rm imperfect}\rangle = \sqrt{1-\tilde\epsilon}\;
|\Psi^{\rm perfect}\rangle + \sqrt{\tilde\epsilon}\;
|\Psi_\perp^{\rm perfect}\rangle \label{evapthm} \;,
\end{equation}
for some perturbation $0\le \tilde\epsilon<\epsilon$. 
The demonstration of robustness to imperfections 
completes our proof of the no-hiding theorem.
{$~$}\hfill \rule{2mm}{2mm}
\vskip 0.1truein

This result comes as a surprise if we consider another, extensively studied,
example of a hiding process~---~state randomization \cite{random}.
There it has been shown that inexact randomization of an arbitrary pure 
state of dimension $d$ can be performed with an ancilla of dimension 
$O(d\log d)$ whereas exact randomization requires an ancilla of dimension 
at least $d^2$. The inexact state-randomization therefore cannot be 
expressed in general as a mere perturbation from the perfect case. 
By enriching the class of states to be hidden to include states which may 
be entangled to some reference system, we have demonstrated robustness, 
with a dimension-independent perturbative degradation.  Indeed, this is 
crucial for any application where the dimensions of the various subspaces 
involved may be unknown and possibly infinite.  

{\bf Teleportation revisited:}
By the no-hiding theorem, the direct quantum analogue of 
the one-time pad is impossible for arbitrary input states. 
Notwithstanding this, quantum states can still be transformed into
pure correlations between three or more subsystems \cite{secret} (not 
counting the external reference subsystem). This underscores 
the analogy between quantum teleportation and the one-time pad. To apply 
the no-hiding theorem to teleportation, we require a globally quantum 
description which we obtain by enlarging the ancilla to include the 
`environment' (or measurement system) \cite{sam1} used to decohere 
Alice's Bell-state. For a single qubit in an arbitrary pure 
state $|\psi\rangle$, the teleportation protocol reduces to
\begin{equation}
|\psi\rangle \label{3evap} 
\rightarrow \frac{1}{2}
\sum_{j,k=0}^{1} |2j+k\rangle_{\rm\strut Alice} \!\!
\otimes |2j+k\rangle_{\rm \strut message} \!\!
\otimes \sigma_z^j\sigma_x^k\; |\psi\rangle_{\rm \strut Bob}\nonumber \;.
\end{equation}
To complete the protocol, Bob need only utilize the value of the
message to undo the randomizing 
operations to retrieve
$|\psi\rangle$. It is easy to check that each of the three subsystems
in Eq.~(\ref{3evap}) is in the maximally mixed state for that space. 
Thus, the information appears only as inter-subsystem 
correlations. (Relabeling the subsystems of  Eq.~(\ref{3evap}) yields an
alternative tri-partite analogue to the one-time pad \cite{Leung}.)
However, the above observation does not contradict the no-hiding theorem.

In fact, our key result can be recovered by rewriting the teleportation 
process in terms of a bi-partite system.  For instance, since the reduced 
density matrix of Bob's subsystem contains no information about the hidden 
state $|\psi\rangle$, it must lie entirely in the remainder of Hilbert space. 
Indeed, it is easy to check that the state $|\psi\rangle$ is completely 
encoded within the union of the Alice and message subsystems \cite{Oppenheim}.
(The same argument holds for Alice's subsystem or for the message 
subsystem.) Hence from a purely quantum mechanical perspective, 
teleportation is consistent with our result. Indeed, this unitary variation 
of the teleportation protocol could serve as an experimental verification 
of the no-hiding theorem, where the bi-partite systems could be reconstructed 
separately via quantum-state tomography to identify in which subsystem 
the original qubit was encoded.

{\bf Thermodynamics:}
The no-hiding theorem offers deep new insights into the nature of 
quantum information. In particular, it generalizes Landauer's erasure 
principle \cite{Landauer}, according to which any process that erases a bit 
of information must dump one bit's worth of entropy into the environment. 
Landauer's principle applies universally to classical or quantum information 
\cite{Landauer}. However, the no-hiding theorem applies to any process
hiding a quantum state, whether by erasure, randomization, thermalization 
or any other procedure. In this sense, quantum information hiding is 
equivalent to its erasure, whereas classical information hiding is 
fundamentally distinct from erasure.

Landauer's principle provides fundamental insight into thermodynamic
reasoning, such as in the resolution of Maxwell's demon. In contrast, data 
hiding provides more insight into the nature of thermalization processes.  
The terminology used above --- input, output and ancilla --- now takes on 
thermodynamic interpretations (e.g., initial system, final system, 
environment; or input system, output radiation, environment, etc.)
In the simplest case of a single system and environment, as the state of
the system thermalizes, it contracts to a thermal distribution independent 
of its initial description. Perfect hiding implies complete thermalization, 
whereas imperfect hiding may shed some light on the approach to an 
equilibrium state.  Either way, as the state vanishes from one subspace,
it must appear in the remainder of Hilbert space (i.e., the environment). 
To apply the no-hiding theorem, we must consider an enlarged purified 
environment, or super-environment. As in teleportation, we again find 
ourselves with three subsystems: system, environment and the remaining 
supra-environment.  We can conclude that the quantum information that 
vanished is appearing somewhere in the complete environment (including 
correlations between the two subsystems therein).

{\bf Black-hole evaporation:}
Having proved the no-hiding theorem in an abstract quantum-information 
theoretic setting, let us now consider its implications for information flow 
in and out of black holes.

Hawking's seminal work on black-hole evaporation some 30 years ago
\cite{swh0} precipitated a crisis in quantum physics.  Hawking's 
calculations showed that whatever matter falls into it, a black hole 
evaporates in a steady stream of ideal featureless 
radiation. In Hawking's semi-classical analysis this radiation is 
completely independent of the in-falling matter, at least until the 
black hole has shrunk to near the Planck mass. For massive black holes 
(many times a Planck mass) Hawking's analysis should presumably be 
arbitrarily good. Nothing in Hawking's semi-classical approach changes
if the black hole were created, or continued to be fed, with matter whose 
quantum states are entangled with external (reference) degrees-of-freedom.
However, in such a scenario, we can immediately apply the no-hiding
theorem. The in-falling matter would correspond to subsystem $I$ and
the out-going Hawking radiation would be subsystem $O$ in our formulation.
Thus, within the framework of Hawking's semi-classical analysis, the
no-hiding theorem implies that no information is carried either within 
the out-going radiation or in correlations between the out-going 
radiation and anything else. This strong rejection of the correlations 
option is based on two assumptions alone: unitarity and Hawking's 
semi-classical analysis of the radiation. 

We stress that the exact nature of the Hawking radiation (e.g., 
whether it is black-body or gray-body \cite{Beck}) is irrelevant to our 
argument --- in particular, it need not be thermal --- so long as the 
reduced state of the outgoing radiation field is independent of the 
detailed state of the in-fallen matter.  Furthermore, we note that the 
state of the in-fallen matter may be subject to a number of superselection 
rules disallowing certain superpositions. In that sense, the in-fallen 
matter is not truly in an arbitrary quantum state. Nonetheless, up to 
that nuance, any subspaces corresponding to the allowed superpositions 
must obey the no-hiding theorem.

We now expose the severity of the black-hole information crisis in one 
specific formulation of the paradox \cite{pageFeed}. Suppose one feeds 
a black hole (with externally entangled states) at the Hawking-emission 
rate for an arbitrarily long time. Then, Hawking's semi-classical 
analysis would predict that such a black hole, of a fixed size, could 
contain an unbounded amount of entropy, associated with the states of the 
in-fallen matter. This unbounded information density is itself tantamount 
to a loss of unitarity (at least in our universe) \cite{JP}. This 
formulation of the black-hole information paradox is particularly 
instructive as it applies to black holes of arbitrary size.

Naturally, one would always expect some deviations from Hawking's analysis. 
For instance, although a tiny effect, there should at least be some small 
scattering of in-falling matter off outgoing Hawking radiation. This is
where robustness is key. If various perturbations lead to deviations 
of size $\epsilon$ from perfect featureless radiation then Eq.~(\ref{evapthm})
quantifies the deviation away from the ideal no-hiding theorem.
Whether this deviation is carrying away information directly or via 
correlations or through interference with the main contribution is 
immaterial; the net amount of information that may be carried away in 
this manner would be $O(\sqrt{\epsilon})$ or more likely $O(\epsilon)$. 
Since these deviations are believed to be vanishingly small for truly 
cosmologically-sized black holes this route to even a partial resolution 
to the black-hole information paradox now appears untenable.

The no-hiding theorem provides new insight into the different laws
governing classical and quantum information. Unlike classical bits, 
arbitrary quantum states cannot completely hide in correlations between 
a pair of subsystems.  A robust statement of this result leads to a severe 
formulation of the black-hole information paradox: Either unitarity fails 
or Hawking's semi-classical predictions must break down. The no-hiding 
theorem rigorously rules out any ``third possibility'' that the information
escapes from the black hole but is nevertheless inaccessible as
it is hidden in correlations between semi-classical Hawking radiation 
and the black hole's internal state. This provides a criterion to test 
any proposed resolution of the paradox: Any resolution that preserves
unitarity must predict a breakdown in Hawking's analysis \cite{swh0} even 
for cosmologically-sized black holes.

\vskip 0.07truein
The authors appreciate discussions with N.\ Cohen, J.\ Eisert,
D.\ Kretschmann, 
T.\ Sudbery, L.\ Vaidman, A.\ Winter and R.\ Werner.
SLB holds a Wolfson - Royal Society research merit award.


\begin{thebibliography}{99} 

\bibitem{Shannon} C.\ E.\ Shannon,
Bell Systems Tech.\ J.\ {\bf 28}, 656 (1949).

\bibitem{swh0} S.\ W.\ Hawking,
Nature {\bf 248}, 30 (1974);
Comm.\ Math.\ Phys.\ {\bf 43}, 199 (1975);
Phys.\ Rev.\ D {\bf 14}, 2460 (1976).

\bibitem{page} D.\ N.\ Page,
Phys.\ Rev.\ Lett.\ {\bf 44}, 301 (1980);
{\bf 71}, 1291 (1993); {\bf 71}, 3743 (1993);
Int.\ J.\ Mod.\ Phys.\ D {\bf 3}, 93 (1994).

\bibitem{thoft} G.\ 't Hooft,
Nucl.\ Phys.\ {\bf B256}, 727 (1985);
Nucl.\ Phys.\ {\bf B335}, 138 (1990).

\bibitem{secret} R.\ Cleve, D.\ Gottesman and H.-K.\ Lo,
Phys.\ Rev.\ Lett.\ {\bf 83}, 648 (1999).

\bibitem{Collins} See, e.g., D.\ Collins and S.\ Popescu,
Phys.\ Rev.\ A {\bf 65}, 032321 (2002).


\bibitem{wil} F.\ Wilczek,
in {\it Black Holes, Membranes, Wormholes, and Superstrings}, 
eds.\ S.\ Kalara and D.\ V.\ Nanopoulos,
(World Scientific, Singapore, 1993), p.\ 1.

\bibitem{schiffer93} U.\ H.\ Danielsson and M.\ Schiffer,
Phys.\ Rev.\ D {\bf 48}, 4779 (1993).

\bibitem{pres94} T.\ M.\ Fiola, J.\ Preskill, A.\ Strominger and
S.\ P.\ Trivedi,
Phys.\ Rev.\ D {\bf 50}, 3987 (1994).

\bibitem{kraus} P.\ Kraus and F.\ Wilczek,
Nucl.\ Phys.\ B {\bf 433}(2), 403 (1995).

\bibitem{bose} S.\ Bose, L.\ Parker and Y.\ Peleg,
Phys.\ Rev.\ D {\bf 54}, 7490 (1996).

\bibitem{obadia} N.\ Obadia and R.\ Parentani,
Phys.\ Rev.\ D {\bf 67}, 024022 (2003).


\bibitem{NC} M.\ Nielsen and I.\ Chuang,
{\it Quantum Computation\/} (CUP, Cambridge, 2000) p.400.

\bibitem{Schu} B.\ Schumacher and M.\ D.\ Westmoreland,
Quantum Information Processing {\bf 1}, 5 (2002).


\bibitem{random} P.\ Hayden, D.\ Leung, P.\ W.\ Shor and A.\ Winter,
Commun.\ Math.\ Phys.\ {\bf 250}, 371 (2004).

\bibitem{sam1} S.\ L.\ Braunstein,
Phys.\ Rev.\ A {\bf 53}, 1900 (1996).



\bibitem{Leung} D.\ W.\ Leung,
Quantum Inf.\ Comput.\ {\bf 2}(1), 14 (2002).

\bibitem{Oppenheim}J.\ Oppenheim and M.\ Horodecki, quant-ph/0306161.

\bibitem{Landauer} R.\ Landauer,
IBM J.\ Res.\ Dev.\ {\bf 3}, 183 (1961).
C.\ H.\ Bennett,
Int.\ J.\ Theor.\ Phys.\ {\bf 21}, 905  (1982).

\bibitem{pageFeed} D.\ N.\ Page, in {\it Abstracts of Contributed
Papers for the Discussion Groups, 9th International Conference on
General Relativity and Gravitation}, ed.\ E.\ Schmutzer,
(Friedrich Schiller University, Jena, 1980), p.\ 582.

\bibitem{JP} J.\ Preskill,
in {\it Black Holes, Membranes, Wormholes, and Superstrings}, 
eds.\ S.\ Kalara and D.\ V.\ Nanopoulos,
(World Scientific, Singapore, 1993), p.\ 22.

\bibitem{Beck} J.\ D.\ Bekenstein,
Phys.\ Rev.\ lett.\ {\bf 70}, 3680 (1993).


\end{thebibliography}
\end{document}